\documentclass[pss]{wiley2sp} 
\usepackage{amsmath}

\begin{document}

\title{Electron transmission through step- and barrier-like potentials in graphene ribbons}

\titlerunning{Electron transmission through step- and barrier-like potentials \ldots}

\author{%
  Yuriy Klymenko\textsuperscript{\textsf{\bfseries 1}}, 
Lyuba Malysheva\textsuperscript{\textsf{\bfseries 2}}, 
and  Alexander Onipko\textsuperscript{\textsf{\bfseries 2,
\Ast}} }

\authorrunning{Yuriy Klymenko et al.}

\mail{e-mail
  \textsf{aleon@ifm.liu.se}, Phone
  +46-13-288904, Fax +46-13-28 8969}

\institute{%
  \textsuperscript{1}\,
Space Research Institute of National Academy of Sciences and National Space Agency of Ukraine, 03187, Kyiv, Ukraine
\\
 \textsuperscript{2}\, Bogolyubov Institute for Theoretical Physics, 03680, Kyiv, Ukraine}

\received{XXXX, revised XXXX, accepted XXXX} 
\published{XXXX} 


\abstract{The list of textbook tunneling formulas is extended by deriving exact expressions for the transmission coefficient in graphene ribbons with armchair edges and the step-like and barrier-like profiles of site energies along the ribbon. These expressions are obtained by matching wave functions at the interfaces between the regions, where quasiparticles have constant but different potential energies. It is shown that for an $U_0$ high barrier and low-energy electrons and holes, the mode transmission of charge carriers in this type of ribbons is described by the textbook formula, where the constant barrier is replaced by an effective, energy-dependent barrier, $U_0\rightarrow U(E)$. For the lowest/highest electron/hole mode, $U(E)$ goes, respectively, to zero and nonzero value in metallic and semiconducting ribbons. This and other peculiarities of through-barrier/step transmission in graphene are discussed and compared with related earlier results.}

\maketitle   

\section{Introduction.}
In recent literature, a considerable attention has been paid to modeling of charge transport in graphene \cite{W1,W2,Fal,Kats1,Kats2,Been,Peres1,M-R,Hw,W3,Rob}.
This report gets in focus two classic problems of Quantum Mechanics which have been discussed in Refs. \cite{Fal,Kats1,Kats2,Been}: Particle transmission in a potential that has a step-like or barrier-like profile. The transmission coefficient has been shown essentially different from the textbook formula \cite{Land}. In particular, the normal incidence of electrons or holes on the interface between two regions with different potential energies results in the full transmission without backscattering \cite{Fal,Kats1}. In the Dirac relativistic quantum mechanics, a similar behavior of the transmission of massless fermions is known as the Klein paradox \cite{Klein}. Because the quoted analytical treatments were based on the use of the Dirac equation and matching either the wave functions or transfer matrices at the interfaces of different regions, the rederivation of the Klein result was natural rather than surprising. As is well known, this approach for graphene is restricted to energies around the point of neutrality that implies that the long wave limit is valid \cite{Ando}. However, it is not at all ensured {\it a priori} that all calculations in this approximation lead to the results which follow from exact calculations and then, passing to the long wave limit. Here, the transmission coefficient is found by exploiting the matching technique, as in Refs. \cite{Kats1,Kats2,Been}, but taking into account the exact energy spectrum of graphene. In other words, we solve the Schr\"odinger equation to which the Dirac equation is a certain approximation. Our general conclusion is that the validity of relativistic approach to the description of charge transport in graphene ribbons is limited, strictly speaking, to the zero-mode transmission in metallic ribbons.

Our model system is an armchair graphene ribbon (GR) with site energies taking zero and $U_0$ values as illustrated in Fig.~1. This choice is dictated by the following considerations. First, armchair GRs can have either metallic or semiconducting spectrum depending on the ribbon width \cite{Nakada}. Thus, the quantum conductance of both basic graphene materials can be studied on equal footings. Second, distinct from zigzag GRs, the armchair GR spectrum does not have a special band of edge states that complicates the description. Third, the long wave limit of the transmission coefficient (obtained here with the account to the discreteness of graphene ribbons) can be compared with earlier derived formulas that makes this analysis particularly instructive.

\begin{figure*}[htb]%
\centering
\includegraphics*[width=0.9\linewidth,height=0.5\linewidth]{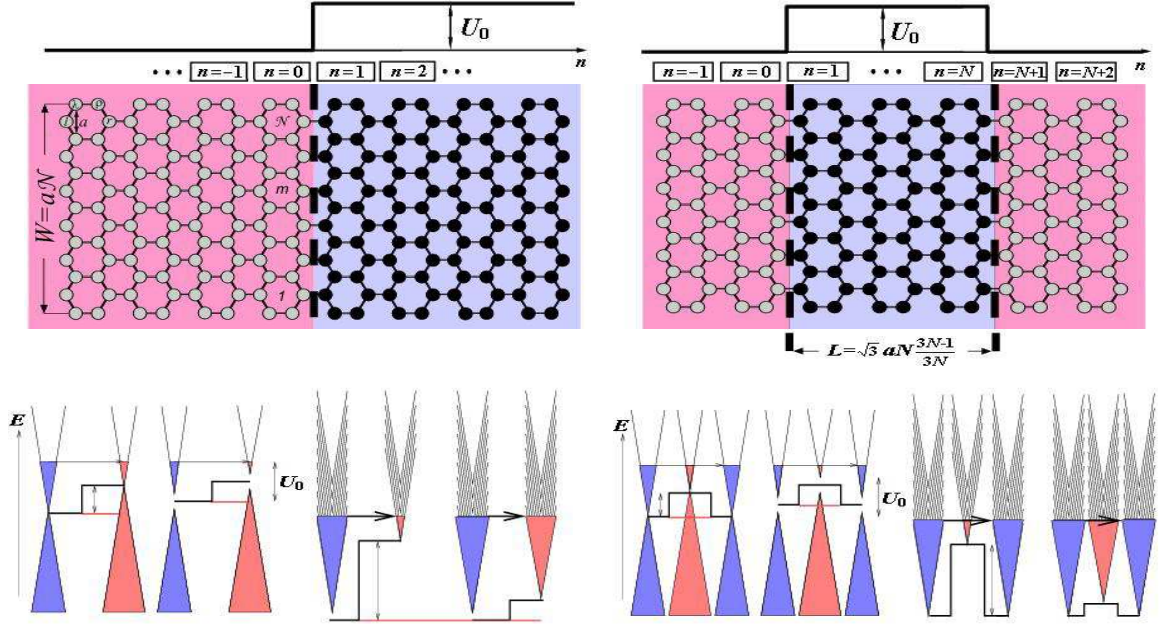}
\caption{ 
Upper part: An armchair graphene ribbon with a step-like (left) and barrier-like (right) profiles of site energies; gray and black circles corresponds to C atoms with site energies equal to zero and $U_0$, respectively (filled by hydrogens dangling bonds along edges are not shown). Blue (red) shaded regions indicate a higher (lower) concentration of electrons. Lower part: Schematic representation of the singlemode and multimode electron transmission in an armchair graphene ribbon. A step-like and barrier-like change of site energies $U_0$ models a gate voltage $ V_{\rm g} $, $U_0=eV_{\rm g}$; red and black horizontal lines correspond to the neutrality point, $E=E_F=0$. On the right, only conduction bands are shown. The potential difference associated with a voltage source that drives electrons from the left to the right is not shown. $a\approx$\,0.246 nm. 
}
\label{Fig2}
\end{figure*}

\section{Transmission coefficient.}
The solution to the stationary Schr\"odinger equation $H\Psi=E\Psi$ with the nearest-neighbor tight-binding Hamiltonian for a honeycomb lattice has the form (C-C hopping integral $|t|$ is a unit of energy)
\begin{equation}\label{1}
\Psi = \sqrt{\frac{2}{{\cal N}+1}} \sum_{m=1}^{{\cal N}+1}\sum_{n=-\infty}^\infty\sum_{\alpha} 
\psi_{m,n,\alpha}^j |m,n,\alpha\rangle,
\end{equation}
where $|m,n,\alpha\rangle$ is the $2p_z$ orbital at the $\alpha$th atom of benzene ring with coordinates $\{m,n\}$, $\alpha=l,r,\lambda,\rho$ with the meaning of labels explained in Fig.~1; $|{\cal N}{\rm +1},n,\alpha{\rm =}l,r\rangle$ = 0. According \cite{Lyuba1}, coefficients $\psi_{m,n,\alpha{\rm =}l,r}^j$=\,$\sin(\xi_jm)\phi_{n,\alpha}^j$, $\xi_j=\pi j/({\cal N}{\rm +}1)$, $j$ = 1,2,...,$\cal N$, and $\phi_{n,\alpha}^j$ satisfies equation (the site energy is zero)
\begin{equation}\label{2}
\phi_{n,\alpha}^j  =g_{\alpha,l}^j\phi_{n-1,r}^j+ g_{\alpha,r}^j \phi_{(n+1),l}^j,
\end{equation}
where $g_{l,r}^j =g_{r,l}^j $, $g_{l,l}^j=g_{r,r}^j $, ${\cal D}_j g_{l,r}^j =4\cos^2(\xi_j/2)$,
${\cal D}_j g_{l,l}^j =E[E^2-1-4\cos^2(\xi_j/2)]$, and zeros of ${\cal D}_j= [E^2-4\cos^2(\xi_j/2)]^2-E^2$ determine the $\pi$ electron spectrum of an $\cal N$-long acene, C$_{4{\cal N}+2}$H$_{2{\cal N}+4}$. For site energies equal to $U_0$, $E\rightarrow \bar E$ = $E-U_0$.

As shown in \cite{Lyuba1,Lyuba3}, states of $\pi$ electrons in armchair ribbons can be classified in 2$\cal N$ "$j$-minus" and "$j$-plus" conduction 1D bands $E_j^\pm(\kappa^\pm_j)$ and equal number of valence bands $-E_j^\pm(\kappa^\pm_j)$.
Since we are interested in the wave-like solutions to the above equation,
 $ \phi_{n,\alpha}^j =\tilde{\phi}_\alpha^j\exp(i\kappa^\pm_jn)$, 
where $\kappa^\pm_j$ satisfies the dispersion relation
\begin{equation}\label{3}
E_j^{\pm\,2}=1\pm4\cos({\xi_j}/{2})\cos (\kappa^\pm_j/2)+4\cos^2({\xi_j}/{2}),
\end{equation}
it is convenient 
to introduce the phase shift between the $l$th and $r$th sites: $\tilde{\phi}_l^j  =\tilde{\phi}_r^je^{i(\theta_j-\kappa^\pm_j)}$, $e^{i\theta_j}=g_{l,l}^j/(1-g_{r,l}^j e^{i\kappa^\pm_j})$. 
It will be seen soon that the change of this phase, $\theta_j$$\rightarrow$$\bar\theta_j$, that corresponds to a change of site energies by $U_0$, $E_j^\pm$$\rightarrow$$E_j^\pm$$-$$U_0$ $\equiv$ $\bar E_j^\pm$ and $\kappa^\pm_j$$\rightarrow$$ \bar\kappa^\pm_j$, plays an important role in determining electron transmission to/through a region, where the site energy equals $U_0$.

With reference to Fig. 1, the wave function (\ref{1}), describing incoming from the left and reflected backward or transmitted to the right electrons, can be represented as
\begin{equation}\label{4}
\phi^j_{n,r}=\left\{
\begin{array}{ll}
e^{i\kappa^\pm_jn}+r_{j}e^{-i\kappa^{\pm} _jn}, & n{\rm <}1,\\
t_{j}e^{i\bar \kappa^{\pm} _jn}, & n{\rm \ge}1,
\end{array}
\right.
\end{equation}
\begin{equation}\label{5}
\phi^j_{n,l}=\left\{
\begin{array}{ll}
e^{i[\kappa^\pm_j(n-1)+\theta_j]}+r_{j}e^{-i[\kappa^\pm_j(n-1)+\theta_j]},& n{\rm < 1,}\\
t_{j}e^{i[\bar \kappa^\pm_j(n-1)+\bar\theta_j]}& n{\rm \ge}1,
\end{array}
\right.
\end{equation}
for the step-like potential, and
\begin{equation}\label{6}
\phi^j_{n,r}=\left\{
\begin{array}{ll}
e^{i\kappa^\pm_jn}+r_{j}e^{-i\kappa^\pm_jn}, & n{\rm <}1,\\
a_je^{i\bar \kappa^\pm_jn}+b_je^{-i\bar \kappa^\pm_jn}, & n\in \overline{1{\rm ,}N},\\
t_{j}e^{i\kappa^\pm_jn}, & n{\rm >}N,
\end{array}
\right.
\end{equation}
\begin{equation}\label{7}
\phi^j_{n,l}=\left\{
\begin{array}{ll}
e^{i[\kappa^\pm_j(n-1)+\theta_j]}+r_{j}e^{-i[\kappa^\pm_j(n-1)+\theta_j]},& n{\rm < 1,}\\
a_je^{i[\bar \kappa^\pm_j(n-1)+\bar\theta_j]}+b_je^{-i[\bar \kappa^\pm_j(n-1)+\bar\theta]}_j,&n \in \overline{1{\rm ,}N},\\
t_{j}e^{i[\kappa^\pm_j(n-1)+\theta_j]}& n{\rm >}N
\end{array}
\right.
\end{equation}
for the barrier-like potential. Matching the wave functions at the interfaces between regions with different site energies gives us the necessary equations for unknown coefficients $a_j$, $b_j$, $r_j$, and $t_j$. Finding the amplitude of transmitted $j$ wave yields the transmission coefficient for the given mode,
\begin{equation}\label{8}
T_j = |t_j|^2 \left \{
\begin{array}{l}
\left|E\sin(\bar \kappa^\pm_j/2)/[\bar E\sin (\kappa^\pm_j/2)]\right|,\\
1,
\end{array} \right.
\end{equation}
where the upper and lower lines refer to the potential step and barrier, respectively. The total transmission coefficient, $T(E)=\sum_jT_j$ is determined by the number of "open" modes, see below.

\subsection{Step $U_0$.}
By exploiting $t_j$ from Eqs.~(\ref{4}) and (\ref{5}), we obtain 
\begin{equation}\label{9}
T_j =\frac{|\sin \theta_j \sin \bar\theta_j|}{\sin^2[(\theta_j\pm \bar\theta_j)/2]},
\end{equation}
or, rewritten as a function of wave vectors,
\begin{equation}\label{10}
T_j =
\frac{\cos^2 (\xi_j/2) \sin (\kappa^\pm_j/2) \sin (\bar \kappa^\pm_j/2) }
{\left|\cos^2 (\xi_j/2)\sin^2[(\kappa^\pm_j{\rm \pm}\bar \kappa^\pm_j)/4]-(U_0/4)^2\right|}.
\end{equation}
The upper (lower) sign in these equations corresponds to $\bar E>0$ ($\bar E<0$); energy is supposed to be positive; due to the spectrum symmetry $T_j(E,U_0)=T_j(-E,-U_0)$.

Note that dispersion relation $\bar E_j^\pm(\bar \kappa^\pm_j)$ can be satisfied by both real and imaginary values of $\bar \kappa^\pm_j$. However, according to Eq. (\ref{5}) (and in analogy with the textbook treatment \cite{Land}) imaginarity of $\kappa^\pm_j$ makes $T_j$ zero. Distinct from the textbook case is that the unit transmission can occur at $\theta_j=\bar\theta_j$; fulfillment of this equality does not necessarily requires $U_0=0$ that is the usual condition of the unit transmission.

Here, our prime interest concerns the energy region close to the Fermi energy $E_F=0$ of undoped graphene, where the long wave approximation, $\kappa^\pm _j,|E|<<1$ (also implying $|U_0|<<1$), provides a reliable description. For this energy region, the energy scale $\sqrt{3}|t|/2$ is more convenient. Henceforth, it is used instead of $|t|$ together with a new notation $\sqrt{3}k_x\equiv\kappa^-_{j=j^*\pm\mu}$ with $\mu$ = 0,1,... $<<\cal N$. The new variable satisfies the following set of dispersion relations,
$$
E= \pm\sqrt{k_i^2 + k_x^2},
$$
\begin{equation} \label{11}
 k_i= \frac{\pi}{{\cal N}{\rm +1}}
\left \{
\begin{array}{ll}
|{\rm \pm}\mu|,& i=\mu,\\
\frac{1}{3} i,&i=1,2,4,5, \dots, 3\mu{\rm -1}, 3\mu{\rm +1}, \dots
\end{array} \right.
\end{equation}
which are equivalent to Eq. (\ref{3}) in the long-wave limit \cite{Note}. 
The upper line of Eq. (\ref{11}) refers to metallic GRs, $j^* =2({\cal N}{\rm +1)/3}$ is an integer; 
 lower line refers to semiconducting GRs, where $j^*=(2{\cal N} +3)/3$ or $j^*=(2{\cal N} +1)/3$ are integers. 

For small energies, Eq. (\ref{10}) takes the form,
\begin{equation}\label{12}
T_i =
\frac{4 k_x\bar k_x }{\left|(k_x\pm\bar k_x )^2-U_0^2\right|}
\end{equation}
which can be rewritten with the use of Eq. (\ref{11}) as
\begin{equation}\label{13}
T_i =2
\frac{\sqrt{ (E^2- k_i^2 )(\bar E^2-k_i^2)}}{\left|\pm\sqrt{(E^2- k_i^2 )(\bar E^2-k_i^2)} +E\bar E -k_i^2 \right|}.
\end{equation}
According to this equation, the zero-mode transmission in metallic GRs has the unit probability, $T_0=1$, independent of the value of $U_0$. This specifies the absence of backscattering under the normal incidence in graphene $n$-$p$ junctions \cite{Fal} in the context of metallic graphene ribbons.

The obtained result is worthwhile comparing with the textbook formula for the probability of over-step transmission, $D=\sqrt{E\bar E}/\left(\sqrt{E}+\sqrt{\bar E}\right)^2$ \cite{Land}. The difference between transmission in metallic and semiconducting GRs is substantial only in the case of singlemode or few mode transmission. For energies $E$$>>$$|U_0|$, that is $E\approx$$\bar E$, $T_i\approx$1 for the mode majority, independent of whether the ribbon has a metallic or semiconducting spectrum. The general behavior of $T$$=$$\sum_{i=0}^{i^{\rm max}}T_i$, $i^{\rm max}>>$1, as a function of $U_0$ is as follows: For $|U_0|=qE$, $q>>1$, $T$$\sim$$2/q$; $T(U_0$=0)$\approx 2i^{\rm max}$; $T(U_0$=$E$)=0; and $T(U_0$=2$E$)=$T_{\rm max}$$<$$2i^{\rm max}$.

\subsection{Barrier $U_0$.}
By finding $t_j$ from Eqs.~(\ref{6}), (\ref{7}) and substituting it in Eq.~(\ref{8}) we obtain
\begin{equation}\label{14}
T_j=\frac{\sin^2\theta_j\sin^2\bar{\theta}_j}{\sin^2\theta_j\sin^2\bar{\theta}_j+
\left(\cos\theta_j-\cos\bar{\theta}_j\right)^2\sin^2(\bar \kappa^\pm_jN)}.
\end{equation}
Similarly to Eq.~(\ref{9}), this representation is characteristic for graphene structures. It shows, in particular,  that the unit transmission occurs under the coincidence of phases $\theta_j$ and $\bar\theta_j$. This can be regarded as a sort of new resonances which differ from the familiar condition of resonances for the over-barrier transmission, $\sin(\bar \kappa^\pm_jN)$ = 0. 

In terms of wave vectors, Eq.~(\ref{14}) has the form 
\begin{equation}\label{15}
T_j=\frac{\sin^2 \kappa^\pm_j\sin^2 \bar \kappa^\pm_j}{\sin^2 \kappa^\pm_j\sin^2
\bar \kappa^\pm_j+U_j^2\sin^2(\bar \kappa^\pm_jN)},
\end{equation}
where (in units of $|t|$)
\begin{equation}\label{16}
\begin{array}{l}
U_j=\\
\displaystyle
\frac{|U_0|}{2}\left| \left[E\bar E{\rm -1+}4\cos^2(\xi_j/2) \right] 
\frac{\cos(\kappa^\pm_j/2)\cos(\bar \kappa^\pm_j/2)}{\cos^2(\xi_j/2)}\right|. 
\end{array}
\end{equation}

To facilitate the comparison with earlier results, it is instructive to look at these expressions for small energies of electrons and holes. In the long-wave limit, an analogue of Eq. (\ref{12}) for the barrier-like potential reads
\begin{equation}\label{17}
T_i=\frac{k_x^2 \bar k_x^2}{k_x^2\bar k_x^2+ k_i^2 U_0^2\sin^2(\sqrt{3}\bar k_x N)}.
\end{equation}
Again, $T_0=1$ for metallic ribbons. Except this case,  under the replacement $k_iU_0\rightarrow U_0/2$ (and the above mentioned convention regarding energy units) the above equation coincides with the textbook formula for the through/over-barrier transmission \cite{Land}. 

If the energy of incident electrons is tuned to the neutrality point of the scattering region, $E=U_0$, it follows from Eq.~(\ref{17})
\begin{equation}\label{18}
T_i=\frac{k_x^2}{k_x^2+U_0^2\sinh^2\left(\sqrt{3}k_i N\right)}. 
\end{equation}
Formula (4) derived in Ref. \cite{Been} for this case reads (in original notations)
\begin{equation}\label{19}
T_n=\frac{1}{\cosh^2Lq_n+(q_n/k_\infty)^2\sinh^2Lq_n},
\end{equation}
where, to our understanding, $k_\infty^2=k_x^2+q_n^2$, with $k_x$ and $q_n=\pi(n+1/2)/W$, $n=0,1,...$, having the meaning of the longitudinal (along the ribbon) and transverse components of the wave vector, respectively. Whatever reading used, we could not agree our Eqs.~(\ref{17}) and Eq.~(\ref{18}) neither with the above equation  
nor with Eq.~(8) from the same reference.

As seen from Eqs.~(\ref{17}) and (\ref{18}), the multimode and singlemode transmissions must be distinguished. In the multimode transmission, $E\approx k_{i^{\rm max}},\, i^{\rm max}>>1$, and for $|E|>>|U_0|$, $T_i(E)\approx 1$ for all open modes, $|\bar E|>k_i$. In this case, the total transmission coefficient for metallic and semiconducting GRs differs only marginally. In contrast, in the case of singlemode transmission, the difference is substantial. Because of the zero and finite values of $k_0$, $T=1$ for metallic GRs, but for semiconducting GRs, $T\sim \exp\left(-\frac{\pi L}{3W}\right)$, if $k_0<|E|<<k_1$ and $L/W>1$. The latter result agrees with the expression for the Green function of $N\times\cal N$ honeycomb lattice \cite{Lyuba4}.

Another formula worth mentioning in the present context refers to an expression for the tunneling probability that was obtained in \cite{Kats1} for massless Dirac fermions with (dimensional) kinetic energy $E=\pm\hbar v_F\sqrt{k_x^2+k_y^2}$ in a 2D space $-\infty<x,y<\infty$, where the potential energy is $U_0$, if $0\le x\le L$, and zero otherwise. Assuming the equivalence of this expression for energy with Eq.~(\ref{11}) and by using the correspondence $\sqrt{3}a|t|/2=\hbar v_F$ and 
$$k_y^2 \leftrightarrow \frac{\pi^2}{W^2}\left \{
\begin{array}{l}
|\pm\mu|^2 ,\\
|\mu\mp \frac{1}{3}|^2 ,
\end{array} \right.
$$
it can be proved that Eq.~(\ref{17}) and an expression for $T$ that follows from Eq.~(3) in Ref. \cite{Kats1} have exactly the same form. 

To wind up this report, we would like to emphasize that the methodology used for the derivation of the new formulas for the transmission coefficient in armchair graphene ribbons is the same as in \cite{Fal,Kats1,Kats2,Been}. Distinct from the previous considerations is the use of exact solutions of the model Hamiltonian for the description of $\pi$ electron states in ideal armchair ribbons. 

The authors acknowledge the support of this work by Visby program of the Swedish Institute (Si).


\begin{thebibliography}{[1]}
\bibitem{Note}
A more accurate approximation reveals the fine structure of the band spectrum of metallic GRs \cite{Lyuba3}. Here, it is disregarded, but its incorporation into the final formulas is streightforward.
\end{thebibliography}
\end{document}